\documentclass[useAMS,usenatbib]{mn2e}
\input psfig.sty

\title[Structure and evolution of low mass close binaries]
{Structure and evolution of low mass W UMa type systems}

\author[Li, L. et al.]
{Lifang Li \thanks{E-mail:
gssephd@public.km.yn.cn or lifang\_li@hotmail.com}, Zhanwen Han and Fenghui
Zhang\\ National Astronomical Observatories/Yunnan Observatory, Chinese
Academy of Sciences, P.O.Box 110,\\
Kunming, Yunnan Province 650011,P. R. China
}

\begin{document}

\date{Accepted yy mm dd. Received yy, mm, dd; in original form 2003 November 5}

\pagerange{\pageref{firstpage}--\pageref{lastpage}} \pubyear{2003}

\maketitle

\label{firstpage}

\begin{abstract}
The structure and evolution of low-mass W UMa type contact binaries
are discussed by employing Eggleton's stellar evolution code
\citep{egg71,egg72,egg73}. Assuming that these systems completely satisfy
Roche geometry, for contact binaries with every kind of mass
ratios (0.02$\sim$1.0), we calculate the relative radii ($R_{1,2}/A$,
where $R_{1,2}$ are the radii of both stars, and $A$ the orbital separation)
of both components of contact binaries in different contact depth between
inner and outer Roche lobes. We obtain
a radius grid of contact binaries, and can ensure the surfaces of two
components lying on an equipotential surface by interpolation using this
radius grid when we follow the evolution of the contact binaries. Serious
uncertainties concern
mainly the transfer of energy in these systems, i.e., it is unclear that how
and where the energy is transferred. We assume that the energy transfer takes
place in the different regions of the common envelope to investigate the
effects of the region of energy transfer on the structure and evolution of
contact binaries. We find that the region of energy transfer has significant
influence on the structure and evolution of contact binaries, and conclude
that the energy transfer may occur in the outermost layers
of the common convective envelope for W-type systems, and this transfer takes
place in the deeper layers of the common envelope for A-type systems.
Meanwhile, if we assume that the energy transfer takes place in the outermost
layers for our model with low total mass, and
find that our model steadily evolves towards a system with a smaller
mass ratio and a deeper envelope, suggesting that some A-type W UMa systems
with low total mass
could be considered as the later evolutionary stages of W-subtype systems,
and that the surface temperature of the secondary excesses that of the
primary during the time when the primary expands rapidly, or the secondary
contracts rapidly, suggesting that W-subtype systems may be caused by expansion
of the primary, or by the contraction of the secondary.

\end{abstract}

\begin{keywords}
stars: binaries: close--stars: mass-loss--stars: rotation--stars: evolution
\end{keywords}

\section{Introduction}

It is probable that more than 50 per cent of all stars are in
binary or multiple systems. An unknown, but possibly large,
percentage of these systems are sufficiently close that sometime
during their lifetime interacting as a result of Roche lobe
overflow (RLOF). The W UMa-type contact binaries are the most
common ones, comprising some 95 per cent of eclipsing variables in
the solar neighborhood \citep{shap48} or one stars in every
1000--2000 in the same spectral range \citep{egg67}. Allowing for
selection effects, W UMa stars may even contribute 1 per cent of
all F and G dwarfs \citep{vant75a}. A more recent discussion of
contact binaries in the solar neighborhood is carried out by
\citet{ruc02}. He considers the complete sample of 32 EW, EB, and
ellipsoidal (ELL) variables with $V<7.5$, and gives the frequency
as 1 per 500 stars with $-0.5<V<5.5$, including a serious estimate
of the uncertainty.

W UMa stars are found both in young and in old galactic clusters
\citep{vant75b}. Among these clusters are: 1. NGC 2602, NGC 6383,
NGC 7235 ($\la 10^{7}$ yr); 2. Pleiades, Coma, Praesepe ($\la
5\times 10^{8}$yr); 3. NGC188, M67 ($\ga 5\times 10^{9}$ yr). For
this reason it seems we should be able to construct zero-age as
well as evolved contact models. Further, there would be a natural
explanation for the existence of W UMa stars in young as well as
in old cluster if zero-age contact models could be shown to evolve
on a nuclear timescale. Structure and evolution of contact
binaries are complex and by no means well understood. Although
theoretical investigation on early-type systems is almost absent,
our discussion will be restricted to late-type systems. The
observational properties \citep{mon81,ruc93} that provide the best
clues to the nature of late-type W UMa systems are that the
systems are of fairly low total mass, and are in shallow contact,
that no equal mass systems exist, that the mass-luminosity
relation is unusual, and that most of the W-type systems are to be
un unevolved or slightly evolved only, and some of them are known
to be unevolved.

\citet{luc68} had the key idea of a common convective envelope in
which the entropy is constant and energy is transferred from the
primary to the secondary in the common convective envelope.
Thermal equilibrium turned out to be
usually impossible. Lucy thermal equilibrium model (violating the
period-colour relation) only in a limited mass region and for
extreme Population I composition ($Z\ga 0.04$). \citet{mos70} and
\citet{whe72} encountered similar difficulties. Indeed, thermal
equilibrium was found to be in conflict with the equal entropy
condition in zero age contact binaries \citep{kah95}.

Renouncing the restriction to thermal equilibrium, \citet{luc76,fla76}, and
\citet{rob77} obtained contact binary solution
evolving in thermal cycles about a state of marginal contact. The solutions
are in agreement with the period-colour relation, but they have bad light
curves because of a large temperature difference between the two components
for a considerable part of the time in a cycle. Some authors encountered the
same difficulty, the so-called light curve paradox, also for systems evolving
without loss of contact. \citet{haz01} pointed out that the present
theory of cyclic contact binaries (thermal relaxation oscillation, TRO) is
not in a position to resolve the light curve paradox. The difficulty of
explaining the EW-type light curves (i.e. light curve paradox) reflects a
basic conflict between the treatment of the internal structure and the
observed properties of the outermost layers.

\citet{kah02a,kah02b} has discussed the structure equations of
contact binaries, and assumed that the energy sources/sinks caused
by the interaction of the components occurs only in the
secondary's/primary's outer layers to obtain the models which are
applied to the typical late-type systems. First he imposed the
restriction that the fractional extent in mass of the
sources/sinks in the layers above the critical surface is the same
in the both components, and found solutions evolving in thermal
cycles. And if the energy transfer is assumed to be sufficiently
effective, loss of contact is avoided. Otherwise, the cycles
consists of a long quiet phase in good contact and a short violent
phase with rapid changes between contact, semi-detached, and
detached configurations. And in both cases a large potential
difference between the surfaces of both components occurs during a
part of the contact phase, that is to say, the system deviates
hydrostatic equilibrium during this time. Meanwhile, the maximum
contact degree is extremely small ($\sim 2\%$) in both cases,
suggesting that late-type contact binaries must be shallow
contact. However, the contact
degree of most of real W UMa systems which were observed to be in
good thermal contact is about $10\%$, only
a little of real late-type contact systems (ER Cep, AO Cam,
Maceroni \& van't Veer, 1996) is observed to be in good thermal
contact with a smaller contact degree ($\sim$ 2 per cent).

In a series of papers, Shu and his colleagues have extended the
unequal entropy model by putting an equal entropy common envelope
on top of interiors with the entropy difference, $\Delta S = 0$,
\citep{shu76, shu79,
shu80,lub77,lub79}. These models  satisfy the light curve constraint but
requires a temperature discontinuity region between the common envelope
and the interior of the secondary component. This so-called DSC model has been
strongly attacked by other investigators \citep{haz78,pap79,smi80} since
it was thought to be violating the second law of thermodynamics.

The structure and evolution of W UMa stars still comprises many
unsolved questions although many progresses have been
achieved. The most difficult problem concerns the energy transfer
between the components, i.e., it is not clear where and how the
energy is transferred. The mechanism causing energy transfer is
still a largely unsolved problem. In addition, we
do not confirm that the transfer occurs in the base, or the
outermost layers, or the whole of common envelope although it seems
probable that the transfer occurs in the common envelope, above the inner Roche
critical surface, where the stars are in good contact.

In present paper, we have calculated a relative-radius grid of contact
binaries with different mass ratios and different contact degrees according
to the Roche potential to ensure the surfaces of two components lying
on the same equipotential by interpolation using this radius grid.
In addition, we assume that the energy transfer occurs in the
different regions of common envelope to investigate the effect of
region of energy transfer on the structure and evolution of
contact binaries. As result of our investigation, we find that the region of
energy transfer has a significant influence on the structure and evolution
of the contact binaries, and that the energy
transfer may take place in the outermost layers of the common
envelope for W-type systems, and this probably takes place in the deeper
layers of the common envelope for A-type systems.

\section{Structure equations and boundary condition}
\subsection{Differential equations}

Eggleton's stellar evolution code used in present work has
considered the effect of stellar rotation on stellar structure.
Let $\omega$ be the angular velocity, $m_{i}$ the mass variable,
$g_{i}$ the effective gravity in component $i$, $\sigma_{{\rm
ex},i}$ the source (when positive) or sink (when negative) of
energy per unit of mass caused by the interaction of the
components. Employing standard notation, the basic differential
equations are the following

\begin{equation}
\frac{\partial{\rm ln} r_{i}}{\partial m_{i}} = \frac{1}{4\pi r_{i}^{3}\varrho_{i}}
\end{equation}

\begin{equation}
\frac{\partial {\rm ln} P_{i}}{\partial m_{i}} = \frac{1}{4\pi r_{i}^{2}P_{i}}g_{i},\ \ \ \ \
(g_{i}=\frac{Gm_{i}}{r_{i}^{2}}(1-\frac{2\omega^{2}r_{i}^{3}}{3Gm_{i}}))
\end{equation}

\begin{equation}
\frac{\partial L_{i}}{\partial m_{i}} = \epsilon_{i}-\epsilon_{\rm \nu}-T_{i}\frac{{\rm D}s_{i}}{{\rm D}t} + \sigma_{{\rm ex},i}
\end{equation}

\begin{equation}
\frac{\partial{\rm ln}T_{i}}{\partial m_{i}} = \frac{\partial{\rm ln} P_{i}}{\partial m
_{i}}\nabla_{i}
\end{equation}
where $\nabla_{i}$\ ($i=1,2$) is the temperature gradients of both
components, and given by

\begin{equation}
\nabla_{i}=\biggl\{\matrix{
                            \nabla_{a,i}\ \ \ \ \ \  (\nabla_{r,i}&>&\nabla_{a,i}),\cr
                            \nabla_{r,i}\ \ \ \ \ \  (\nabla_{r,i}&<&\nabla_{a,i}),\cr
                          }
\end{equation}
where $\nabla_{r,a}$ are the radiative and adiabatic temperature
gradients. The radiative temperature gradient is given by

\begin{equation}
\nabla_{r,i}=\frac{3}{16\pi acG}\frac{\kappa_{i}L_{i}P_{i}}{m_{i}T_{i}^{4}}(1-
\frac{2\omega^{2}r_{i}^{3}}{3Gm_{i}})^{-1},
\end{equation}

\subsection{Contact condition}

In hydrostatic equilibrium the surface of each star should coincide with
the same equipotential surface which, in Roche approximation, requires

\begin{equation}
\frac{R_{2}}{R_{1}}=(\frac{M_{2}}{M_{1}})^\beta,
\end{equation}
where $R_{1,2}$ are the radii of both components, $M_{1,2}$ the masses of both
components, $\beta$ the index of mass-radius relation of contact binaries.
The value of $\beta$ varies not only with mass ratio, but also with the depth of contact. It covers a range of $0.45 \sim 0.50$ for the marginal contact
binaries, and will cover a larger range if the contact depth varies.
But most of investigators takes $\beta=0.46$ approximately, which is very close to a
value of a marginal contact binary with a mass ratio of $q\approx 0.5$.
Since the mass ratio
of each contact binary varies during its evolution, some
investigators use a boundary condition as the following
\begin{equation}
\frac{R_{1}}{R_{\rm crit1}}=\frac{R_{2}}{R_{\rm crit2}},
\end{equation}
where $R_{1,2}$ are radii of two components, $R_{\rm crit1,2}$ the Roche
critical radii.
Eqs. (7) and (8) only approximately ensure the surfaces of both components
lying on the same equipotential surface. In order to let the surfaces of
the two stars lying on the same
equipotential accurately, we calculate a radius grid of contact binaries which
have different fixed mass ratios ($0.02,0.04,0.06,\cdot\cdot\cdot,1.0$) and
different contact degrees ($0\%,2.5\%,5\%,\cdot\cdot\cdot,100\%$) for
each fixed mass ratio using a dimensionless Roche potential
\begin{equation}
\Phi=\frac{1}{r_{1}}+\frac{q}{r_{2}} +\frac{1+q}{2}[(x-\frac{q}{1+q})^2+y^2]
-\frac{q^2}{2(1+q)}
\end{equation}
where $\Phi$ is the potential at an arbitrary point P, $r_{1,2}$
the distances of point P from the centers of gravity of the two
stars, $x$ and $y$ the Cartesian coordinates of point P (the
origin of the rectangular system of Cartesian coordinates at the
center of gravity of mass $M_{1}$, and the $x$-axis of which
coincides with the line joining the centers of the two stars, the
$y$-axis in the orbital plane), $r_{1,2}$, $x$, and $y$ are in
unit of orbital separation of the binary,
$q(=\frac{M_{2}}{M_{1}})$ the mass ratio. We can accurately ensure
the surfaces of two stars lying on the same equipotential by
interpolation using our radius grid. Eggleton's stellar evolution
code uses the most modern physics as reported by \citet{pol95},
and it can calculate the evolution of a single star or a binary
(not contact). For evolution of a single star, just choose an
orbital period so large that there is no prospect of RLOF. For
evolution of a detached binary, the evolution of each star is
similar to that of a single star. For evolution of a semi-detached
binary, RLOF has been treated within the code, so that the mass
above Roche lobe is transferred to its companion, and the mass
lost at a rate proportional to the cube of the fractional excess
of the star's radius over its lobe's radius ($\frac{{\rm d}m}{{\rm
d}t}=- C\cdot{\rm Max}[0,({\rm ln}\frac{r}{R_{\rm crit}})^3]$,
where ${\rm d}m/{\rm d}t$ gives the rate at which the mass of the
star changes, $r$ the radius of the star, and $R_{\rm crit}$ the
radius of its Roche lobe), it has been tested throughly and works
very reliably. With $C=1000M_{\rm \odot}$/yr, RLOF proceeds
steadily \citep{han02}.

When a close binary
evolves into a contact one, the mass which is transferred is no
longer the mass above its Roche lobe. If the surface of star 2 is at a
higher potential than star 1, so that an unbalanced pressure gradient
will force mass transferred from star 2 to star 1, then this boundary
condition should be modified as

\begin{equation}
\frac{{\rm d}m_{2}}{{\rm d}t}=-{\rm C}\cdot {\rm max}[0,({\rm ln}\frac{R_{2}}{
R_{20}})^3],
\end{equation}
in which
\begin{equation}
R_{20}=\Re_{2} \root 3 \of {\frac{\varrho_{1}R_{1}^3+\varrho_{2}R_{2}^3}
{\varrho_{1}R_{1}^3+\varrho_{2}\Re_{2}^3}},
\end{equation}
where $\varrho_{1,2}$ are the surface densities of both components,
$R_{1}$ the radius of star 1 (known), $\Re_{2}$ the radius of an equipotential
surface in star2, whose potential is equal to the surface potential of star 1,
and it is obtained by the interpolation according to our radius
grid. $R_{2}$ the real radius of star 2 because of its evolution.
(The mass which is transferred to star 1 is the mass above $R_{20}$ rather
than that above $\Re_{2}$, because the surface potential of star 1 will be higher than that of star 2 if the mass above $\Re_{2}$ is transferred to Star
1. ) Assuming that the radii of the two stars 1 and 2 become $R_{10}$ and
$R_{20}$ respectively after mass transfer, they should satisfy a very accurate
condition ($\frac{R_{10}}{R_1}=\frac{R_{20}}{\Re_2}$). Meanwhile, the mass
gained by star 1 equals to that lost by star 2, i.e., $(\frac{4\pi R_{10}^3}{3}-\frac{4\pi R_{1}^3}{3})\varrho_1=(\frac{4\pi R_{2}^3}{3}-\frac{4\pi R_{20}^3}{3})\varrho_2$. Then Eq. (11) can be easily derived from the two conditions
mentioned above. Eq. (10) gives a mass loss rate from star 2 to star 1. If
the mass is lost by star 1, gained by star 2, the mass loss rate can be
derived in the same way.

For interacting binaries, we put the two stars `side by side' in
the computer, so that 22 differential equations (11 equations
should be solved for each stars), in 22 variables, have to be
solved. The direct coupling between the sets of equations for the
two components is via the boundary conditions and energy transfer
discussed in sect. 3.

\section{A model for Energy transfer}

\citet{str48} first recognized the unusual mass-luminosity
relationship of the secondary components of W Ursae Majoris
systems, suggesting that it might be causally related to a
possible common envelope. \citet{osa64} noted that von Zeipel's
theorem would require the observed approximate constancy of
radiative flux over the surface of a system with a radiative
common envelope in order to maintain hydrostatic equilibrium. The
fact that most W UMa systems appear to have convective envelopes,
however, led \citet{luc68} to propose that the over-luminosity of
the secondaries is directly attributable to energy transfer between the two
stars within a convective envelope.

It is not clear that where and how the energy is transferred from the primary
(more massive star) to the secondary. It seems probable that the transfer
occurs in the common envelope, above the inner Roche lobe, where the stars
are in good contact, but we could not confirm that the energy is transferred
in the base or the outermost layers or the whole of the common convective
envelope. Meanwhile, the mechanism causing energy transfer between the two
components remains uncertain. \citet{mos74,mos76} and
\citet{mos976} have argued that the large-scale circulation envisaged
by \citet{haz73} or \citet{nar76} is likely to be
destroyed by both normal, vertical convection and coriolis forces. They
consider that a more appropriate picture would involve horizontally moving
eddies, again driven by horizontal pressure gradients, which would travel only
a short distant horizontally before of convection; it differs fundamentally
from the mechanism of Hazlehurst \& Meyer-Hofmeister in that the eddies are not
in thermal equilibrium with the surroundings, and will therefore dissipate on a
thermal time-scale even if no other process acts to break them up. On the
contrary, \citet{web77} argues that a large-scale circulation can be
maintained, and has returned to the model of Hazlehurst \& Meyer-Hofmeister.
Therefore, the mechanism causing energy transfer is still largely an unsolved
problem. Because of these uncertainties, most numerical models of contact
binaries have been phenomenological, simply inserting an artificial
energy source
$\Delta L$ in the secondary, usually in the adiabatic part of the common
envelope, and a corresponding energy sink in the primary, the value of
$\Delta L$ being chosen to satisfy the requirements that the system be both
in contact and in equilibrium. As \citet{rob77}, let $L_{1}$ and
$L_{2}$ be the luminosities (nuclear plus thermal luminosities), and $m_{1}$
and $m_{2}$ the masses of the
components. Although the W UMa-type systems are relatively much more
common, $\sim$ 1 per 10 in the equivalent magnitude range, we still assume that the effect of fully efficient energy exchange
is to equalize the light-to-mass ratio of the stars, since this seems to obtain
approximately among the observed systems. If
$\Delta L_{0}$ is lost by the primary and gained by the secondary, we require
that

\begin{equation}
 \frac{L_{1}-\Delta L_{0}}{m_{1}}=\frac{L_{2}+\Delta L_{0}}{m_{2}},
\end{equation}
since transfer is not fully efficient at all phases, an arbitrary factor
$f$ is introduced, which varies through the cycle and goes to zero with
the depth of contact. Thus we take:
\begin{equation}
\Delta L = f\cdot\Delta L_{0},\ \ \  (0\leq f\leq 1).
\end{equation}
where $f$ is the efficient factor of energy transfer. We take
\begin{equation}
f={\rm Min}[1,\alpha(d^{2}-1)]
\end{equation}
in which
\begin{equation}
d={\rm Max}[1,{\rm Min}(\frac{r_{1}}{R_{\rm crit1}},\frac{r_{2}}
{R_{\rm crit2}})]
\end{equation}
where $r_{1,2}$ are the radii of both stars, $R_{\rm crit1,2}$ the
Roche critical radii of both stars. The parameter $\alpha$ is
expected to be moderately large, so that heat transfer becomes
fully efficient for stellar radii exceeding the Roche radii by
some standard small amount.

The luminosity transferred by circulation currents from the
primary to the secondary adopted by \citet{kah02b} is
\begin{equation}
\Delta L = \int_0^{M_{2}} \sigma_{\rm ex,2} {\rm d}m_{2}
= -\int_0^{M_{1}}\sigma_{\rm ex,1}{\rm d}m_{1},
\end{equation}

\section{Evolution into contact}

\begin{figure}
\centerline{\psfig{figure=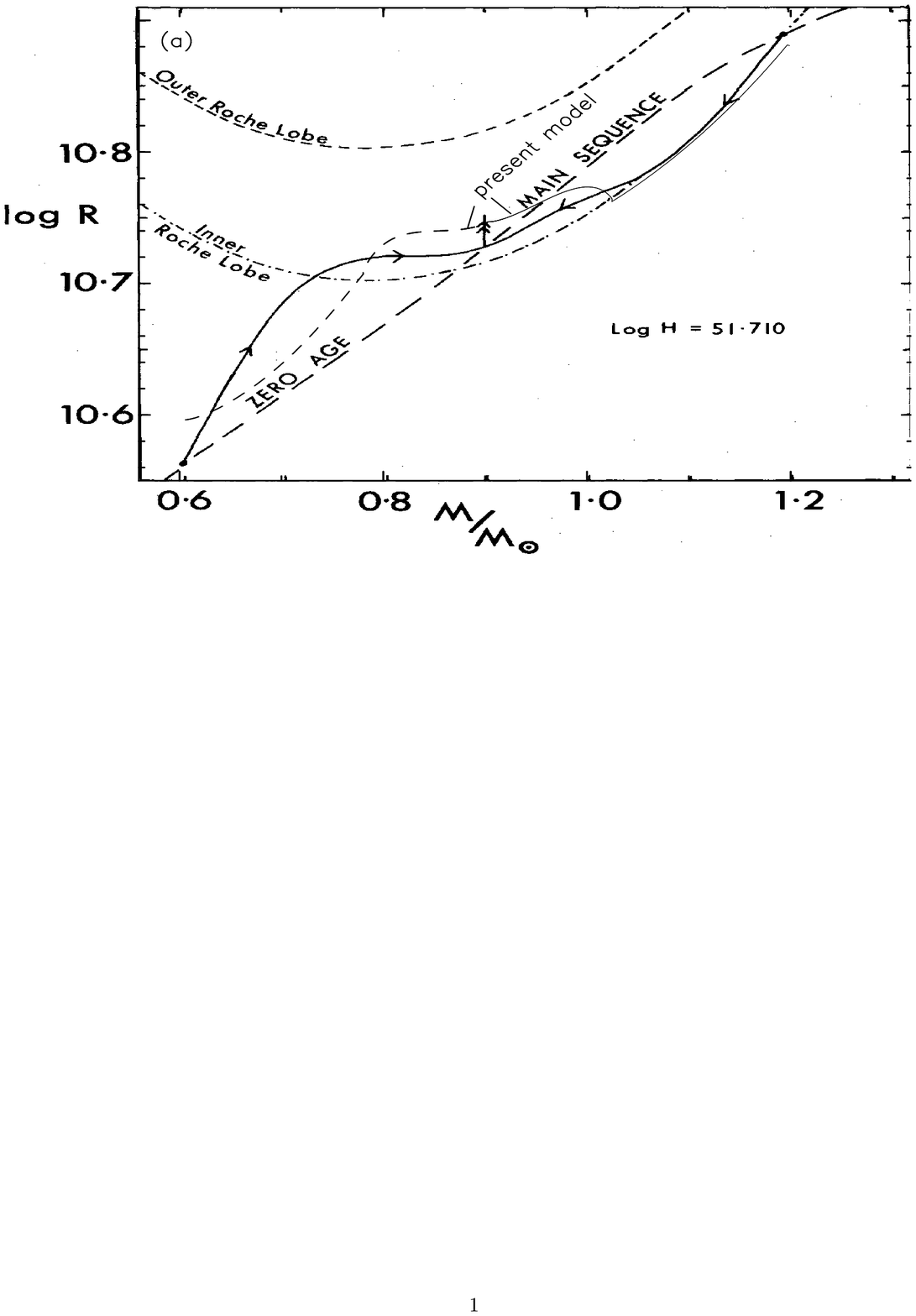,width=8cm,bbllx=72pt,bblly=477pt,bburx=523pt,bbury=754pt,angle=0}} \caption{
Evolution of a close binary system into contact with ${\rm
log}J=51.710$. The primary, starting at the upper right, reached
its Roche lobe almost immediately after leaving the zero-age
main-sequence. Accretion on to the secondary, starting from the
lower left, caused it to swell until it also filled its lobe. The
resulting contact model, or a very similar model, is the initial
model for most subsequent calculations. Further evolution in the
figure assumes (unrealistically) that no heat transfer take place
during contact. The system evolves to equal masses. The present model
represents our result, the others obtained by \citet{rob77}.}
\label{fig1}
\end{figure}
\begin{figure}
\centerline{\psfig{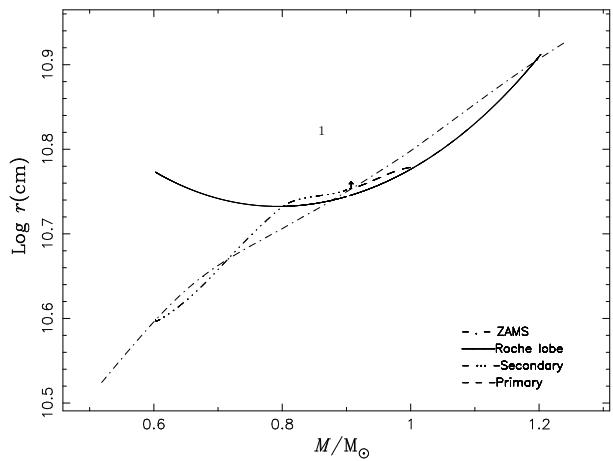}} \caption{
Evolution of a close binary system into contact with ${\rm
log}J=51.723$. }
\label{fig2}
\end{figure}
\subsection{The initial model}

Our initial model consists of two zero-age main-sequence (ZAMS)
stars of Population I ($X=0.70, Z=0.02$) with mass 1.2 and
0.6$M_{\rm \odot}$ as \cite{rob77}. In order to find the influence
of the adoption of our boundary condition and the most modern
physical inputs reported by \citet{pol95} on the evolution of the
binary. At first, we do not consider the rotation of the stars and
the energy transfer between the two components, and take an
initial period of 0.3418 days, and orbital angular momentum, $J$,
of $5.129 \times 10^{51}$ erg$\cdot$s (${\rm log}J$ = 51.710) as
\cite{rob77}. Our result together with Robertson \& Eggleton's one
is shown in Figure 1. It is seen in Figure 1 that the adoption of
our boundary condition and the most modern physical inputs has a
significant influence on the evolution of the binaries. Then we
consider the rotation of both components of the binary, but the
energy transfer between the two components, and take a orbital
period of 0.3732 days, and orbital angular momentum, $J$, of
$5.283 \times 10^{51}$ erg$\cdot$s (${\rm log}J = 51.723$). The
initial separation is about 2.65$R_{\rm \odot}$. The initial model
is a detached binary, and the surface of the primary lies only a
short way inside its Roche lobe which it fills after $2.8\times
10^6$ yr of nuclear evolution. Thereafter it losses mass to its
companion at a rather slow rate which rises approximately $\sim
1.4\times 10^{-8}M_{\rm \odot}/{\rm yr}$. At the beginning of the
mass transfer, the mass loss rate is very slow, the radius of the
secondary is smaller than that of a ZAMS star with the same mass
because the secondary with a convective envelope gains mass and
its radius decreases. When the mass loss rate rises to a higher
value, the secondary is too late to adjust its structure thermally
after it gains mass from the primary and its radius increases so
that the radius of the secondary is larger than that of a ZAMS
star with the same mass (See Figure 2).

The addition mass on to the secondary causes it to expand and its
effective temperature and luminosity to increase, and after a
total of $3.5\times 10^{7}$ yr of evolution from the main sequence
it has swollen to fill its Roche lobe, so that the system evolve
into a contact system.  The contact binary formed at this point
has masses of 1.0 and 0.8$M_{\rm \odot}$ and a mass ratio of 0.8.
Since we do not consider the energy transfer between the two components,
the two stars evolve steadily towards equal masses on a thermal timescale
(see Figure 2). Although some of W UMa-type systems (at least V348 Car,
Hilditch \& Bell, 1987 ) with closely equal masses are consistent with
this scenario, however most of the real W UMa-type systems are not
equal-mass ones. This can be regarded as sufficient proof that the
majority of contact binaries never approach the
equal-mass state, suggesting that the transfer of significant
amount of energy through the common envelope surrounding the
components in direction from the primary to the secondary has the
effect of preventing mass transfer so that most of the real binaries do not
evolve towards contact systems with equal masses.
Therefore, the following discussions only restrict to the possible
consequence of energy transfer.

\subsection{Contact Evolution With Energy Transfer}
\subsubsection{The regions of energy transfer}
\begin{figure}
\centerline{\psfig{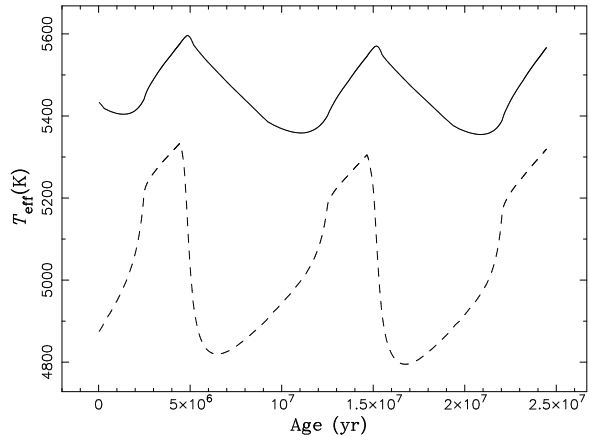}}
\caption{
  The time dependence of the effective temperature of the primary (solid line)
  and the secondary (dashed line) during the first three cycles if energy
  transfer takes place in the base of common envelope.
}
\label{fig3}
\end{figure}

\begin{figure}
\centerline{\psfig{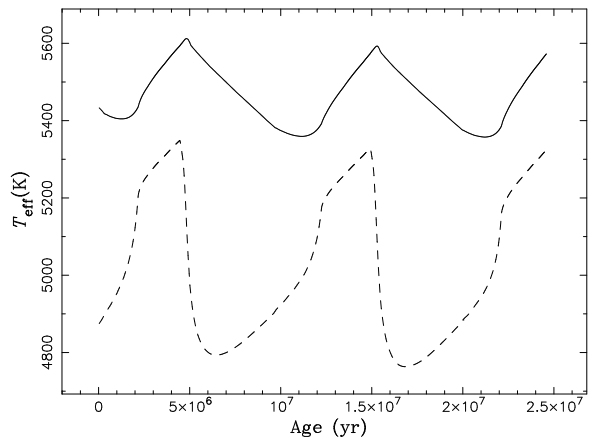}}
\caption{
  The time dependence of the effective temperature of the primary (solid line)
  and the secondary (dashed line) during the first three cycles if energy
  transfer takes place in the whole of common envelope.
}
\label{fig4}
\end{figure}

\begin{figure}
\centerline{\psfig{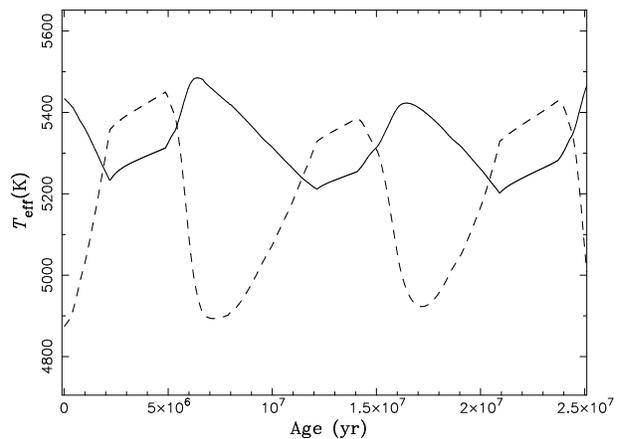}}
\caption{
  The time dependence of the effective temperature of the primary (solid line)
  and the secondary (dashed line ) during the first three cycles if energy
  transfer takes place in the outermost layers of common envelope.
}
\label{fig5}
\end{figure}

We consider now the evolution of the binary system from the
initial contact model with energy transfer included in the manner
described in sect. 3. Nuclear evolution is included in the models, although
the influence of the composition changes is negligible in the early
stage. The differential equations describing both the structure and the
chemical composition for both stars are solved simultaneously.
The luminosity increment $\Delta L$ is
applied to each component with appropriate sign in a relevant
region of both components. As the uncertainty of the region of the
energy transfer, we take $\alpha = 45$, and assume that the energy
transfer takes place in the different regions of the common
envelope to investigate the effect of the energy
transfer region on the structure and evolution of the contact systems. At first,
we assume that
the energy transfer takes place in the adiabatic part in the base
of common convective envelope, and energy increment is applied in
10 meshpoints just above the Roche lobe, so that more
stellar material takes part in energy transfer, and the significant energy
($\sim 0.5L_{\rm \odot}$) can be transferred between the two components.
Then we assume that the energy transfer takes place in the whole
of the common envelope surrounding the two components, and energy
increment is applied in all meshpoints in the common
envelope. Finally, we assume that the energy transfer takes place
in the outermost layers in the common envelope of the two stars,
and the luminosity increment is applied in 10 meshpoints
near the surface of common envelope. The evolution of
the surface effective temperature of both components are plotted
in Figures 3, 4, and 5, respectively. As seen from Figs. 3 and 4,
The same result is obtained when the energy
transfer is assumed to take place in the base or the whole of the common
envelope. In these cases, there is no possibility of the effective temperature
of the secondary (less massive star) exceeding that of the primary at any
time of a cycle even though energy transfer is fully efficient,
and a large temperature difference between the two components
occurs in a considerable time of a cycle when the energy transfer
takes place in the base or the whole of the common envelope. This
would usually be interpreted as implying that these contact models
represent the A-subtype W UMa systems only, but the W-subtype W
UMa systems. It is seen in
Figure 5 that the temperature of the secondary can excess that of
the primary for a considerable time of a cycle, and that a small
temperature difference ($\Delta T<$300K) between the two components occurs
in a large part of the time of a cycle. In this case, the model represents
the structure and evolution of a W-subtype W UMa system during a
considerable time of a cycle when the energy transfer takes place
in the outermost layers (10 meshpoints) near the surface of the
common envelope. Therefore, the region of the energy transfer has a significant
influence on the structure and evolution of the contact binaries. The
same results are obtained if the energy transfer is
assumed to take place in the base or in the whole of the common envelope,
because most of energy is still transferred in the base of the common
envelope due to a higher density of the base of the common envelope
even if the energy transfer is assumed to take place in the whole of the
common envelope. However, it is very different from the result based on the
assumption that the energy transfer in the outermost layers.
Therefore, we conclude that the energy
transfer in W-type systems may take place in  the outermost
layers near the surface of common envelope, and energy transfer in A-type
systems probably takes place in the deeper layers of their common envelope.
\citet{mos76} has proposed a small-scale eddy model underlying the energy
transfer which is similar to mixing-length theory, and
concluded that the energy transfer in contact binaries must
be characterized by a scale of a complicated eddy structure which is much
shorter than the separation between the two stars, and that it appears possible
to divide the envelope into a surface layer where convective mixing
determines the scale and a deeper layer where Coriolis effects dominate.
Meanwhile, each A-type system has a radiative envelope, and the energy transfer
is caused by the eddies due to Coriolis effects, and should take place in the
deeper layers (base) of the common envelope; However each W-type system has
a convective envelope, the energy transfer is attributed to eddies due
to convection, and should take place in the surface layers of the common
envelope. Since the surfaces of both stars are radiative rather than
convective, it follows that convection is by no means essential to heat
transport in contact envelopes, although it may well have an important
influence. In our model with low total mass, we assume the energy transfer
takes place in the outermost layers of the common envelope. The thermal
structure of the two components of
the binary (${\rm log} J =51.723, \alpha= 45$) during a phase of contact
evolution and of semi-detached evolution is shown in Figure 6. In
contact phase, energy transfer takes place in the outermost layers of
the common envelope, and the outermost layers of the
envelopes of both components are so similar that the difference is slight.

It is seen in Figure 6a that the luminosity of the secondary's outer
layers below the energy sources is very low, and in these layers
the core luminosity (includes thermal and nuclear luminosities) has almost completely been exhausted by
negative values of the $\epsilon_{g}$-terms in the energy balance,
i.e. by the expansion of the secondary \citep{kah02b}.

\subsubsection{Cyclic evolution}

\begin{figure*}
\centerline{\psfig{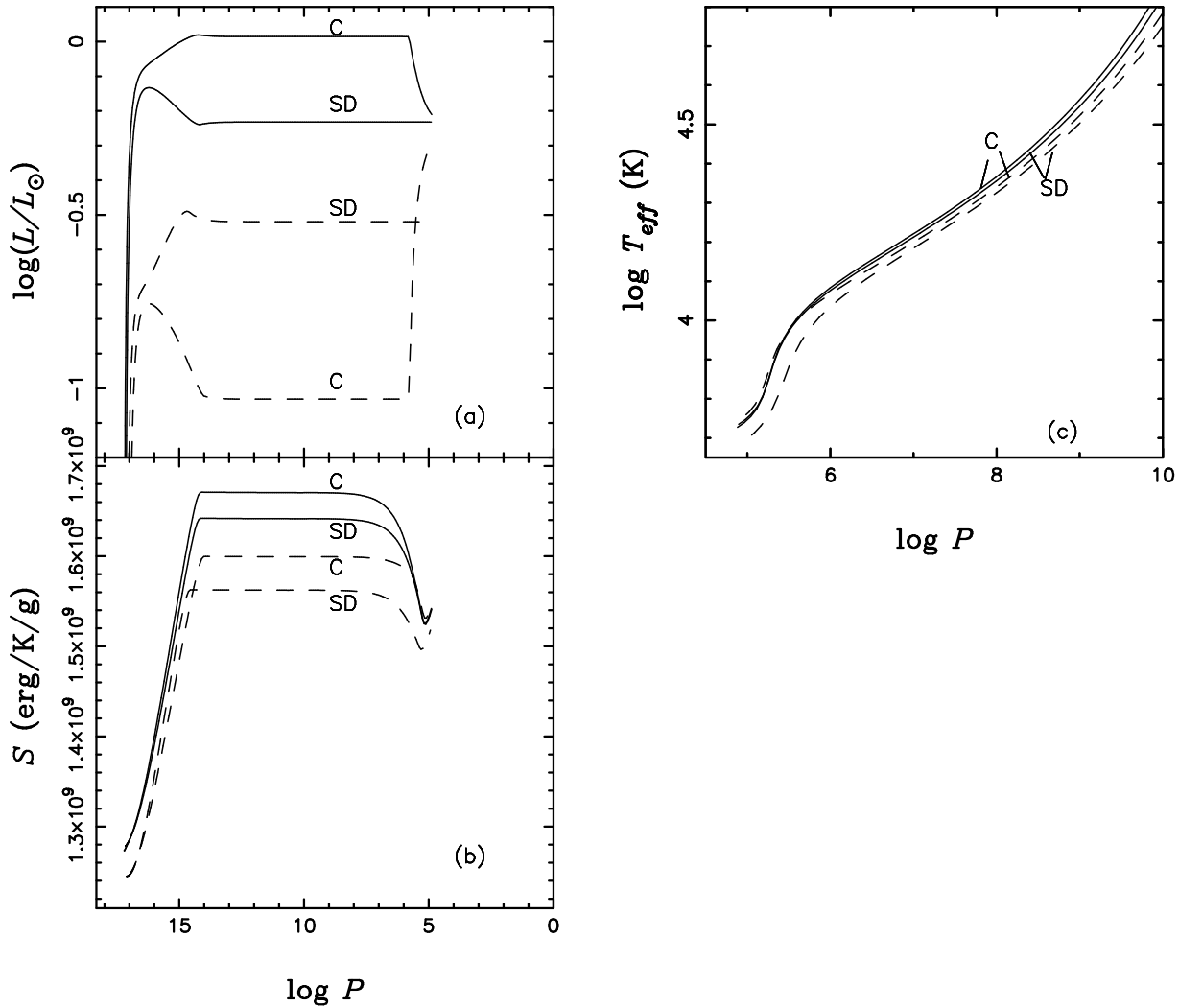}}
\caption{
  The thermal structure of the primary (solid line) and the secondary (dashed line) during a phase of semidetached
  evolution (SD) and of contact evolution (C) with efficient energy transfer.
  In contact, luminosity transfer takes place in the outermost layers of common
  envelope.
}
\label{fig6}
\end{figure*}

We consider now the evolutionary behavior of the binary, also beginning
at the contact model with a mass ratio of about 0.8. The evolution of the binary, with energy
transfer included and the luminosity increment is applied in the
outermost  layers (10 meshpoints) of the common envelope, is shown in Figures
7, 8 and 9 for
$\alpha = 45$.  The system undergoes thermal cycles on a thermal
timescale. At the early stage, the evolution of the contact phase
resemble the properties in the case of no energy transfer, since
the efficiency of energy transfer is extremely low and equilibrium
configuration remains at the equal-mass state. However, the
expansion of the secondary is hastened by the energy which is
being deposited in its envelope, and the growth in the depth of
the common envelope which these causes lead to the continued
increase of energy transfer. The added mass onto the
secondary stops when the significant energy is transferred from
the primary to the secondary. This followed by a rapid rise in the
rate of mass transfer back to the primary, and the rate of mass transfer
rises rapidly to a higher value because of the rapid increase in the heat
transfer, then decreases rapidly to stable value (see Figure 8c), however
these could not prevent a further increase in the depth of contact
and in the luminosity transfer (see Figure 8d and Figure 9c)
until the decrease in radius of the secondary caused by mass loss
can not be compensated by its increase caused by luminosity
transfer.
If this has occurred, the binary evolves rather slowly towards smaller mass
ratios as each star attempts to obtain thermal equilibrium.
Throughout this phase the secondary contracts rapidly and the depth of
contact decreases so that this phase will be rapidly terminated when full
efficient energy transfer can no longer be maintained. This phase
of good thermal contact is characterized by a remarkable constancy
of most of properties of both stars.

\begin{figure}
\centerline{\psfig{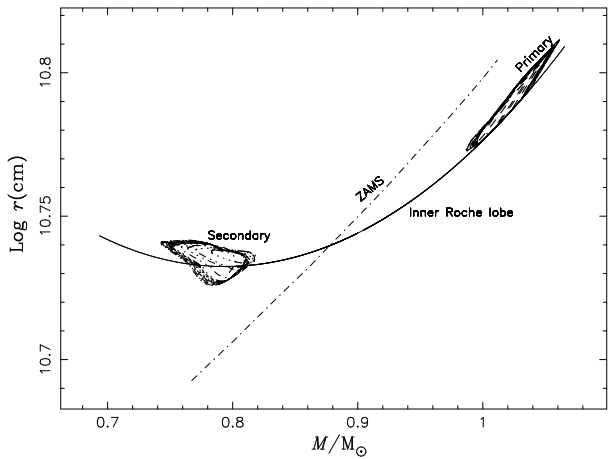}}
\caption{
  The cyclic evolution in the mass-radius diagram of a binary (${\rm log} J=51.723$ and $\alpha= 45$) with alternating contact and semi-detached phases. A steady cycle appears to be set up after two or three initial loops. Heat transfer is
assumed to be fully efficient only when the depth of contact exceeds a certain fraction.
}
\label{fig7}
\end{figure}

Once full efficiency is lost, the secondary, no longer adequately
supported, contracts very rapidly and the direction of mass
transfer reverses, the rate of mass transfer rising to its highest
value in the cycle. The secondary breaks contact with its Roche
lobe, and continues to collapse towards a main-sequence
equilibrium state, its temperature and luminosity falling rapidly.
The luminosity of the primary also decreases during semi-detached
phase, since the Roche lobe again contracts the star to prevent
its free expansion towards thermal equilibrium, and the process of
raising matter up through the star, for transfer to the secondary,
requires significant quantities of energy, at the expense of the
surface luminosity. Meanwhile, the luminosity produced by nuclear
sources also falls because of the core of the primary is also
expanding and cooling during semidetached phase \citep{rob177}. Since
the thermal timescales of the two components are not equal,
the radius of the secondary will increase
before the secondary collapses to be a main-sequence equilibrium state
during semi-detached phase, so that the system evolves into a contact binary
again. Therefore, the two stars of the system are unlikely in thermal
equilibrium at any time of a cycle.

The radius-mass diagram is shown in Figure 7. It is seen in Figure 7 that
the radius of the secondary is larger than that of a ZAMS star with the
same mass, and the radius of the primary is smaller than that of a ZAMS
star with the same mass even though the system evolves in semidetached
phase. It suggests that the two components of the system are not
in thermal equilibrium at any time of a cycle. As seen from Figures 8 and 9,
the system
undergoes cyclic evolution on a thermal timescale, with a period of about
$10^{7}$yr, and the system spends a large part of the time of a
cycle in contact evolution (lasting about 70 percent of the time),
only a small part of time in semidetached evolution. Meanwhile, in
a long phase (lasting a bout 60 percent of the time of a cycle),
the system is in good thermal contact with a temperature
difference between the two components not larger than 300 K.
\citet{kah02b} had given a model in which the temperature difference between
two components
is less than 300K during 80 percent time of a cycle, but the maximum
contact degree of his model is extremely low (about 2\%) which is much lower than
the mean contact degree of the real
W-type systems of about 13\% \citep{smi84}, suggesting that
his model could not evolve in high contact depth. However, our model
can evolve in a maximum contact degree of about 7\% (see Figure 8d) which
is very close to the mean value of W-type systems. The
evolution of the binary in any cycle does not completely repeat
the evolutionary track of the previous cycle, although the
evolution of the binary undergoes the thermal cycles. As seen from
the Figure 8a, the mean mass of the primary in any cycle is larger
than that in the previous cycle, so that the mean mass ratio of the
system becomes smaller and smaller along with the evolution of the binary.
It indicates that the system will evolve to a typical W-type system
with a mass ratio of about 0.5 although a contact system which
originates from a initially detached or semi-detached binary has
a higher mass ratio of about 0.7 \citep{mon81}.
Meanwhile, the maximum contact depth of the binary becomes higher
and higher because of the evolution of the system (see Figure 8d).
The evolution of the luminosities of the two stars are not synchronous,
also are the radii of them, i.e., the luminosity (radius) of the secondary
does not rise to the maximum when the luminosity (radius) of the
primary reaches the minimum, and vice versa. It is caused by the
different thermal timescales of both components. Because
the thermal timescales of both components are unequal, the two
stars unlikely reach the thermal equilibrium, but the two components
of the system attempt to reach the thermal equilibrium, and the attempt
of the system to reach the nonexistent thermal equilibrium, coupled with
Roche geometry, is the driver for the cycling behavior \citep{rah81}.
The evolution of
the temperatures and the orbital period of the two components of the system
is shown in Figure 9b,d. It is seen in Figure 9 that the
mean period in any cycle of the system becomes longer and longer
as the evolution of the binary, and that the temperature of the
secondary excesses that of the primary in a large part of the time
in any cycle.

\begin{figure*}
\centerline{\psfig{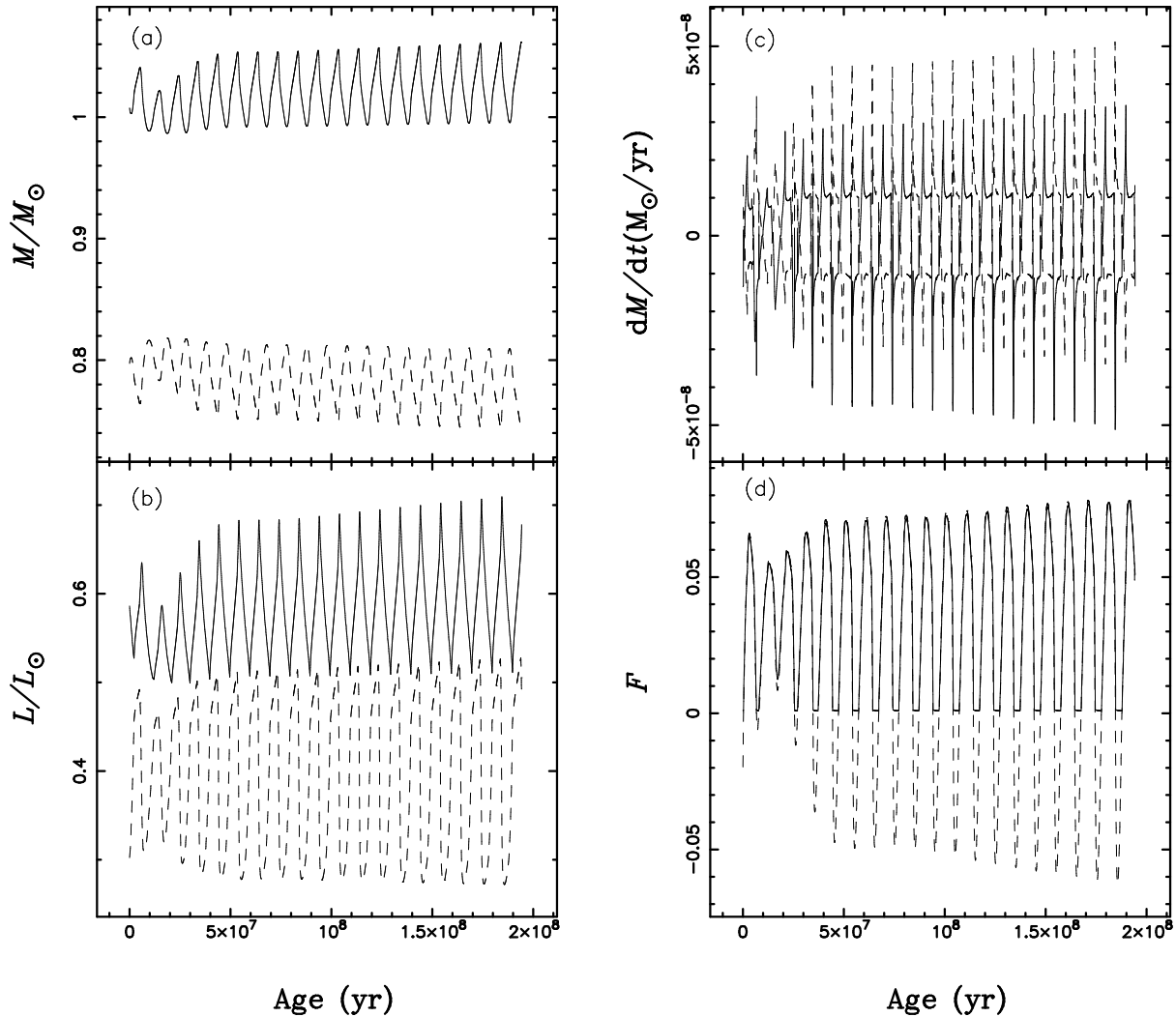}}
\caption{
  The evolution of some quantities [such as the masses, radii, luminosities, mass loss rates, and contact degrees ($F=\frac{\Omega-\Omega_{\rm in}}{\Omega_{\rm out}-\Omega_{\rm in}}$)] of the primary (solid lines) and the secondary (dashed lines ) of the binary as a function of Age, A steady cycle appears to
be set up after two or three initial loops.
}
\label{fig8}
\end{figure*}

\begin{figure*}
\centerline{\psfig{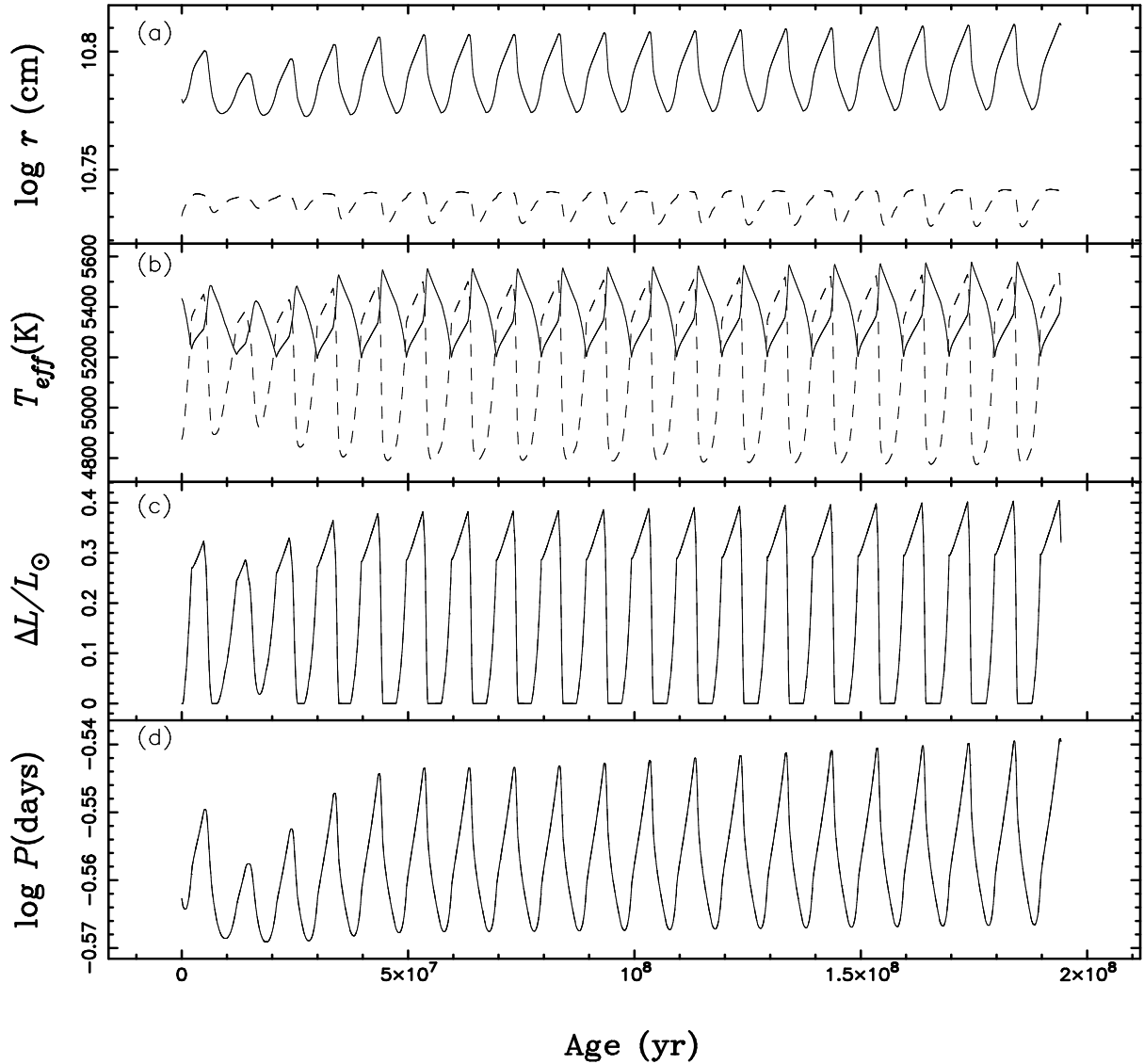}}
\caption{
   The evolution of the radii and temperatures of the
  primary (solid line) and the secondary (dashed line), and also
  the orbital period (in days) and the transferred luminosity of the system.
}
\label{fig9}
\end{figure*}

\subsubsection{The cycles in the period-colour diagram}

\begin{figure}
\centerline{\psfig{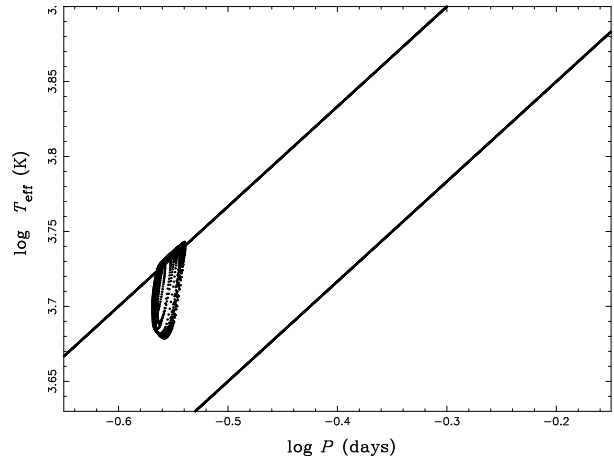}}
\caption{
  The cycles of the less massive component of the binary in the period-colour diagram.
}
\label{fig10}
\end{figure}

As shown by \citet{egg61,egg67}, the observed W UMa systems are located
in a strip of the period-colour diagram, the two boundaries of the period-
colour diagram are written by \citet{kah02b} as
\begin{equation}
1.5{\rm log} T_{\rm eff} - {\rm log} P=5.975...6.15,
\end{equation}
where $P$ is the period in days, which is limited by the solid lines in
Figure 10. Also shown is the cycles of the less massive component of
our model (${\rm log}J=51.723$, $\alpha=45$). As seen from the Figure
9, our model is in agreement with the period-colour relation. Our result
is different from the results obtained by \citet{fla76} and \citet{rob77}, and
similar to that derived by \citet{kah02b}.
The difference between our result and those derived by \citet{fla76} and
\citet{rob77}
is mainly caused by the effect of the stellar rotation considered by us.

\section{Discussion and Summary}

The most uncertainties in the theoretical investigation of contact binaries
concerns mainly the energy transfer, i.e., it is not clear that where and
how this energy is transferred from the primary (more massive star) to the
secondary. Although it seems probable that the transfer occurs in the common
envelope, above the Roche lobe, where the two stars are in good contact.
However, we can not confirm that the energy transfer occurs in the base or the
outermost region or the whole of the common envelope.
Therefore, most of results of the previous investigators are based on the assumption that the
energy transfer between the components occurs in the adiabatic
part of the common convective envelope (adiabatic transfer). The
modification due to superadiabatic transfer for W-type systems is small
\citep{whe72,bie73}. Extreme superadiabatic transfer \citep{mos73}
gives more freedom, but \citet{haz74} showed that this case must
be excluded because the heat capacity of the secondary's
subphotospheric layers is too small. In order to investigate the effect of the
region of energy transfer on the structure and evolution of contact binaries,
we assume the energy transfer takes place in the different regions of the
common envelope,
and find that the region of energy transfer
has significant influence on the evolution of the contact binaries.
There is no possibility of the temperature of the secondary exceeding
that of the primary at any time of a cycle if the energy transfer occurs
in the base or the whole of the common envelope, and the temperature of the
secondary can excess that of the primary if the energy transfer takes place
in the outermost layers of the common envelope, suggesting that
during contact phase of the cycle the models are reasonably good agreement
with the observed properties of A-type W UMa systems when the energy transfer
is assumed to take place in the base or the whole of common envelope, and the
model is very consistent with the observed properties of W-type systems when
the energy transfer is assumed to take place in the outermost
layers of the common envelope.
Therefore, we conclude that the energy transfer
may take place in the base of the common envelope for A-type systems, and the
transfer takes place in the outermost layers for W-type systems.
Since the surfaces of both components are radiative rather than convective,
it follows that convection is by no means essential to heat transport in
contact envelopes, although it may well have an important influence.

The W UMa systems have been divided by \citet{bin70} into A-type
and W-type systems according as the primary minimum in the light
curve is a transit, in which the smaller star partially eclipses
the larger, or an occultation, when the larger star is in front.
The observations of W UMa stars have been reviewed by
\citet{mon81}. The main differences between these subclasses are
as the followings: the A-type systems have longer periods, are
hotter, smaller mass ratio, and in better contact (i.e. the higher
contact depth). Meanwhile, \citet{wil78} has found that the A-type
systems are likely evolved stars. For
a low total mass contact binary, we assumed that the energy transfer takes
place in the outermost layers of the common envelope.
Our model indicates that the mass ratio of the binary becomes smaller and
smaller, and the contact depth becomes higher and higher as the evolution
of the system, that is to say, the system steadily evolves towards a
contact binary with a smaller mass ratio and a deeper common envelope. It
suggests that some A-type systems with low total mass
could be considered as later evolutionary stages of W-subtypes, although
W-type systems show greater activity in the form of period and light
curve changes, which were originally thought to occur on a thermal timescale
\citep{ruc74} but now their periods appear to consists of abrupt changes
typically in a timescale of a few months; And A-type systems
possess more stable light curves and less rapidly varying periods.
However, in this scenario the system evolution was towards smaller
mass ratio, longer periods and deeper contact, and the better physical
contact will lead to the gradual disappearance of the W-type
peculiarities (hotter secondary, light curve perturbations, and
frequent period changes). Therefore, some of low total mass systems
can be regarded as the later evolutionary stages of W-type systems.

In order to obtain a model of W-subtype W UMa system in which the temperature
of the less massive star is higher than that of the massive one, the
investigators assumed that the energy transfer takes place in the
superadiabatic part or even in extreme superadiabatic region of the
envelope of the secondary. However \citet{haz74} suggested that the extreme
superadiabatic transfer must be excluded because the heat capacity of the
secondary's subphotospheric layers is too small. In our model, we do not
assume that the energy occurs in the superadiabatic or extreme superadiabatic
part of the envelope of the secondary, but the surface temperature
of the secondary can excess that of the
primary during the time when the radius of the primary increases rapidly,
or the radius of the secondary (less massive component) decreases rapidly
(see figure 9a,b). The temperature of the secondary exceeding that of the primary in our model can be attributed to exhaustion of a part of the nuclear
luminosity of the primary due to the expansion of the
primary, or the release of the gravitation energy of the secondary because of
its contraction. Meanwhile, it suggests that
the two subtype W UMa systems are probably caused by the energy exchange
between the gravitational (potential) energy and thermal one. \citet{wan94}
has showed that the two types of contact binaries are in two different TRO
states: the less massive component of W-types are shrinking whereas the less
massive star of A-types are swelling. The contraction
of the secondary in W-type systems releases some of its gravitational energy,
therefore, it makes the effective temperature of the secondary higher than
the primary. However, In our model, the contraction of the secondary is one of
the mechanisms which can make the temperature of the secondary higher than the primary, but it only
lasts a very short period, so the surface temperature of the secondary
higher than the primary is mainly caused by the depletion of a part of
the luminosity of the primary due to its rapid expansion.

Our models for contact binary systems exhibit cyclic behavior
about a state of marginal contact, with a period of $10^7$ yr. In
cyclic evolution, the two components of the system are unlikely in
thermal equilibrium because of the difference in their thermal
timescales, but they attempt to reach the thermal equilibrium. The
attempt of the system to reach an nonexistent thermal equilibrium,
coupled with Roche geometry, is driver for cycling behavior
\citep{rah81}. A larger temperature difference ($\Delta T_{\rm
eff}> 300 {\rm K}$) between the two components occurs in a part of
the time of a cycle (lasting about 30$\sim$35 percent time of a
cycle). Almost all of the previous investigators thought that this
requires there to be as many short-period binary with EB light
curves as with EW light curves, and that the models for contact
binaries encounter a difficulty, so called light curve paradox.
\citet{ruc02} gives 13 EWs and 5 EBs (and 14 ELLs, which have too
small an amplitude to be classified as EWs or EBs). It is
reasonable to identify the EWs as contact binaries and the EBs as
semi-detached. The ratio of 5/13 is not much out of line with TRO
theory, so we are not sure there is in fact any light curve
paradox. However, the W UMa systems indeed undergo angular momentum loss
without doubt, and the most likely angular momentum loss mechanism
is magnetic braking \citep{hua66,mes68}. The {\it Einstein} X-ray
observations and {\it IUE} ultraviolet observations \citep{eat83}
showed that W UMa systems are strong sources, suggesting surface
activity of the kind we observe on the Sun, and so the presence of
the magnetic fields. The stellar wind would cause magnetic
braking, and we will discuss the evolution of W UMa systems
included the angular momentum loss in our future work. Meanwhile,
we do not consider the energy source at secondary's atmosphere
provided by the accreting matter from the primary at our present
work. This energy source can hasten the expansion of the
secondary, and shorten the time spent in the semi-detached
evolution. We refer to a forthcoming work included this extra
energy source for the secondary.

\section*{ACKNOWLEDGEMENTS}
We acknowledge the generous support provided by the Chinese
National Science Foundation (Grant No. 10273020, 10303006 and
19925312), the Foundation of Chinese Academy of Sciences
(KJCX2-SW-T06), and from Yunnan Natural Science Foundation (Grant
No. 2002A0020Q). We are gratefully acknowledge stimulating
discussions with Prof. R. Q. Huang and Dr. C. A. Tout. Meanwhile
we are gratefully to Prof. Eggleton for his valuable suggestions
which improve the paper greatly.

\bsp

\label{lastpage}

\end{document}